\documentclass[11pt]{article}

\PassOptionsToPackage{nosort}{cite}

\usepackage{graphicx}
\usepackage{float}
\usepackage{chet}
\usepackage{scalerel}
\usepackage{amssymb}
\usepackage{dsfont}
\usepackage{mathrsfs}

\newcommand{\lsp}{\hspace{1pt}}
\newcommand{\llsp}{\hspace{0.5pt}}
\newcommand{\lnsp}{\hspace{-1pt}}
\newcommand{\llnsp}{\hspace{-0.5pt}}
\newcommand{\veps}{\varepsilon}
\newcommand{\cO}{\mathcal{O}}

\renewcommand{\ge}{\geqslant}

\newcommand{\MN}{\ensuremath{M\lnsp\llnsp N}}
\newcommand{\hP}{{\widehat{P}}}
\newcommand{\hd}{{\hat{d}}}

\DeclareMathOperator{\ord}{ord}

\definecolor{darkblue}{rgb}{0.1,0.1,0.7}
\hypersetup{colorlinks,
           linkcolor={darkblue},
           citecolor={darkblue},
           urlcolor={darkblue}
}

\date{April 2019}

\title{Bootstrapping MN and Tetragonal CFTs\\\vspace{5pt}
in Three Dimensions}

\author{Andreas Stergiou}

\affiliation{Theoretical Division, MS B285, Los Alamos National Laboratory,
Los Alamos, NM 87545, USA}

\abstract{Conformal field theories (CFTs) with MN and tetragonal global
symmetry in $d=2+1$ dimensions are relevant for structural,
antiferromagnetic and helimagnetic phase transitions in a wide class of
materials. The study of these theories with the nonperturbative numerical
conformal bootstrap is initiated in this work.  Bounds for operator
dimensions are obtained and they are found to possess sharp kinks in the MN
case, suggesting the existence of full-fledged CFTs.  In the tetragonal
case, no new kinks are found. Estimates for critical exponents are provided
for a few cases describing phase transitions in various materials. In two
particular MN cases, corresponding to theories with global symmetry groups
$O(2)^2\rtimes S_2$ and $O(2)^3\rtimes S_3$, a second kink is found.  In
the $O(2)^2\rtimes S_2$ case it is argued to be saturated by a CFT that
belongs to a new universality class relevant for the structural phase
transition of NbO$_2$ and paramagnetic-helimagnetic transitions of the
rare-earth metals Ho and Dy.  In the $O(2)^3\rtimes S_3$ case it is
suggested that the CFT that saturates the second kink belongs to a new
universality class relevant for the paramagnetic-antiferromagnetic phase
transition of the rare-earth metal Nd.}

\begin{document}

\maketitle

\toc

\newsec{Introduction and discussion of results}
In recent years it has become clear that the numerical conformal bootstrap
as conceived in \cite{Rattazzi:2008pe}\foot{See \cite{Poland:2018epd} for a
recent review.} is an indispensable tool in our quest to understand and
classify conformal field theories (CFTs). Its power has already been
showcased in the 3D Ising~\cite{ElShowk:2012ht, El-Showk:2014dwa,
Kos:2014bka} and $O(N)$ models~\cite{Kos:2013tga, Kos:2015mba,
Kos:2016ysd}, and recently it has suggested the existence of a new cubic
universality class in 3D, referred to as $C_3$ or
Platonic~\cite{Stergiou:2018gjj, Kousvos:2018rhl}. Now that the method has
showed its strength, it is time for it to be applied to the plethora of
examples of CFTs in $d=3$ suggested by the $\veps=4-d$
expansion~\cite{Wilson:1973jj, Pelissetto:2000ek, Osborn:2017ucf}.  This is
of obvious importance, for the bootstrap gives us nonperturbative
information that is useful both for comparing with experiments as well as
in testing the validity of field theory methods such as the $\veps$
expansion in the $\veps\to1$ limit.

In this work we apply the numerical conformal bootstrap to CFTs with global
symmetry groups that are semidirect products of the form $K^n\rtimes S_n$,
where $K$ is either $O(m)$ or the dihedral group $D_4$ of eight elements,
i.e.\ the group of symmetries of the square. These cases have been analyzed
in detail with the $\veps$ expansion and other field theory methods due to
their importance for structural, antiferromagnetic and helimagnetic phase
transitions. This provides ample motivation for their study with the
bootstrap, with the hope of resolving some of the controversies in the
literature.

One of the cases we analyze in detail in this work is that of
$O(2)^2\rtimes S_2$ symmetry. Such theories are relevant for frustrated
models with noncollinear order---see~\cite[Sec.~11.5]{Pelissetto:2000ek},
\cite{Kawamura_1998} and references therein. Monte Carlo simulations as
well as the $\veps$ expansion and the fixed-dimension expansion have been
used in the literature. Disagreements both in experimental as well as
theoretical results described in~\cite{Kawamura_1998}
and~\cite{Delamotte:2003dw} paint a rather disconcerting picture. In this
work we observe a clear kink in a certain operator dimension bound---see
Fig.~\ref{fig:Delta_X_MN_diag} below.  Following standard intuition, we
attribute this kink to the presence of a CFT with $O(2)^2\rtimes S_2$
symmetry. Using existing results in the literature,
namely~\cite{Kos:2013tga}, we can exclude the possibility that this kink is
saturated by the CFT of two decoupled $O(2)$ models. Obtaining the spectrum
on the kink as explained in~\cite{ElShowk:2012hu}, we are able to provide
estimates for the critical exponents $\beta$ and $\nu$ that are frequently
quoted in the literature.\foot{In terms of the dimensions of the order
parameter $\phi$ and the leading scalar singlet $S$ it is
$\beta=\Delta_\phi/(3-\Delta_S)$ and $\nu=1/(3-\Delta_S)$.} We find
\eqn{\beta=0.293(3)\,,\qquad\nu=0.566(6)\,.}[STAcrit]
These results are incompatible with the $\veps$ expansion at order
$\veps^4$, which gives $\beta\approx0.370$ and
$\nu\approx0.715$~\cite[Table II]{Mudrov2}.  Experimental results for
$\beta$ for XY stacked triangular antiferromagnets and the helimagnet
(spiral magnet) Tb are slightly lower, and for the helimagnets Ho and Dy
higher. Based on the results summarized in \cite[Tables I and
II]{Delamotte:2003dw} and \cite[Table 37]{Pelissetto:2000ek} we may
estimate that experimentally $\beta=0.24(2)$, $\nu=0.55(5)$ for XY stacked
triangular antiferromagnets, $\beta=0.23(4)$, $\nu=0.53(4)$ for Tb, and
$\beta=0.39(4)$, $\nu=0.57(4)$ for Ho and Dy. Also, our result for $\beta$
is below the value measured in the structural phase transition of NbO$_2$,
$\beta=0.40^{+0.04}_{-0.07}$~\cite{Pynn_1976}.

Another case of interest is that of CFTs with $O(2)^3\rtimes S_3$ symmetry.
Here we again find a kink---see Fig.~\ref{fig:Delta_X_MN_nondiag}
below---and for the CFT that saturates it we obtain, with a spectrum
analysis,
\eqn{\beta=0.301(3)\,,\qquad\nu=0.581(6)\,.}[Anticrit]
Just like in the previous paragraph, we do not find good agreement with
results of the $\veps$ expansion, where $\beta\approx0.363$ and
$\nu\approx0.702$~\cite[Table II]{Mudrov2}. A CFT with $O(2)^3\rtimes S_3$
symmetry is supposed to describe the antiferromagnetic phase transition of
Nd~\cite{Mukamel0, Mukamel1, Mukamel2, Mukamel3, Bak:1976zz}, but the
experimental result for $\beta$ in \cite{BakLebech}, namely
$\beta=0.36(2)$, is incompatible with our $\beta$ in \Anticrit.

In both $O(2)^2\rtimes S_2$ and $O(2)^3\rtimes S_3$ cases we just
discussed, we find that the stability of our theory, as measured by the
scaling dimension of the next-to-leading scalar singlet, $S'$, is not in
question.\foot{This relies on a spectrum analysis, a procedure explained in
\cite{ElShowk:2012hu} and \cite[Sec.\ 3.2]{Stergiou:2018gjj}.} More
specifically, in both cases the scaling dimension of $S'$ is slightly below
four, while marginality is of course at three. The $\veps$ expansion
predicts that $S'$, an operator quartic in $\phi$, has dimension slightly
above three~\cite[Table I]{Mudrov2}. The purported closeness of the
dimension of $S'$ to three according to the $\veps$ expansion has
contributed to controversies in the literature regarding the nature of the
stable fixed point, with arguments that it may be that of decoupled $O(2)$
models. In fact, there is a nonperturbative argument that the theory of $n$
decoupled $O(2)$ models is stable under the deformation to the $MN_{2,n}$
theory. On the other hand, the bootstrap suggests that the
fully-interacting $O(2)^2\rtimes S_2$ and $O(2)^3\rtimes S_3$ CFTs are also
stable. We elaborate more on these issues at the end of
section~\ref{secMNAnis}.

It is not clear from our discussion so far that our bootstrap bounds are
saturated by CFTs predicted by the $\veps$ expansion. Our bootstrap results
suggest that there is a well-defined large-$m$ expansion in $O(m)^n\rtimes
S_n$ theories. This was verified by the authors of~\cite{Osborn:2017ucf}
for the fully interacting $O(m)^n\rtimes S_n$ theory of the $\veps$
expansion---see v7 of \cite{Osborn:2017ucf} on the
\href{https://arxiv.org/abs/1707.06165v7}{\texttt{arXiv}}. The important
point here is that since the large-$m$ results of the $\veps$-expansion
theory reproduce the behavior we see in our bootstrap bounds, we conclude
that the kinks we observe are indeed due to the theory predicted by the
$\veps$ expansion. We stress that this argument is robust at large $m$, but
may fail at small $m$, e.g.\ $m=2,3$.

As we alluded to above, experimental results for phase transitions in the
helimagnets Ho and Dy as well as the structural phase transition of NbO$_2$
differ from those in XY stacked triangular antiferromagnets and the
helimagnet Tb~\cite{Delamotte:2003dw}. Prompted by these disagreements, we
have explored theories with $O(2)^2\rtimes S_2$ symmetry in a larger region
of parameter space. The idea is that perhaps XY stacked triangular
antiferromagnets and the helimagnet Tb are not in
the same universality class as NbO$_{2}$ and the helimagnets Ho and Dy,
although at criticality both these theories have $O(2)^2\rtimes S_2$ global
symmetry.  We find support for this suggestion due to a second kink in our
bound and a second local minimum in the central charge---see
Figs.~\ref{fig:Delta_X_MN-2-2} and \ref{fig:cc_MN-2-2} below.  Although
this kink is not as sharp as the one described above, a spectrum analysis
yields
\eqn{\beta=0.355(5)\,,\qquad\nu=0.576(8)\,.}[heliCrit]
These numbers are in good agreement with experiments on
paramagnetic-helimagnetic transitions in Ho and Dy, $\beta=0.39(4)$,
$\nu=0.57(4)$~\cite[Table II]{Delamotte:2003dw}, \cite[Table
37]{Pelissetto:2000ek} and with the structural phase transition of NbO$_2$,
where $\beta=0.40^{+0.04}_{-0.07}$~\cite{Pynn_1976}.

The critical exponent $\beta$ in \Anticrit is not in good agreement with
that measured for the antiferromagnetic phase transition of Nd
in~\cite{BakLebech}. Exploration of theories with $O(2)^3\rtimes S_3$
global symmetry in a larger part of the parameter space reveals a second
kink and a second local minimum in the central charge, much like in the
$O(2)^2\rtimes S_2$ case---see Figs.~\ref{fig:Delta_X_MN-2-3} and
\ref{fig:cc_MN-2-3}  below. At the second kink we find
\eqn{\beta=0.394(5)\,,\qquad\nu=0.590(8)\,,}[secCFT]
in good agreement with the measurement $\beta=0.36(2)$ of~\cite{BakLebech}.

The result of our analysis is that there exist two CFTs with $O(2)^2\rtimes
S_2$ symmetry and two CFTs with $O(2)^3\rtimes S_3$ symmetry. In the
$O(2)^2\rtimes S_2$ case, the first CFT, with critical exponents given in
\STAcrit, is relevant for XY stacked triangular antiferromagnets and the
helimagnet Tb.  The second, with critical exponents given in \heliCrit, is
relevant for  the structural phase transition of NbO$_2$ and the
helimagnets Ho and Dy. In the case of $O(2)^3\rtimes S_3$ symmetry, we only
found experimental determination of the critical exponent $\beta$ in Nd in
the literature~\cite{BakLebech}. It agrees very well with the exponent
in~\secCFT, computed for the CFT that saturates the second kink. We should
mention here that all CFTs appear to have only one relevant scalar singlet,
which in experiments would correspond to the temperature. The $\veps$
expansion finds only one CFT in each case, and does not appear to compute
the critical exponents and the eigenvalues of the stability matrix with
satisfactory accuracy.

Our conclusions do not agree with the suggestion of~\cite{Tissier:2001uk,
Delamotte:2003dw} that in frustrated systems the phase transitions are of
weakly first order. The reason for this is that we find kinks in our
bootstrap bounds and we suggest that they arise due to the presence of
second-order phase transitions.  Note that our determinations of the
correlation-length critical exponent $\nu$ are in remarkable agreement with
experiments. In some cases mild tension exists between our results for the
order-parameter critical exponent $\beta$ and the corresponding
experimental measurements.

For CFTs with symmetry $D_4{\!}^n\rtimes S_n$ we have not managed to obtain
any bounds with features not previously found in the literature or not
corresponding to a symmetry enhancement to $O(2)^n\rtimes S_n$. The $\veps$
expansion does not find a fixed point with $D_4{\!}^n\rtimes S_n$ symmetry.
The lack of kinks in our plots combined with the lack of CFTs with such
symmetry in the $\veps$ expansion suggests that they do not exist in $d=3$.
However,  bootstrap studies of $D_4{\!}^n\rtimes S_n$ CFTs in larger
regions of parameter space are necessary before any final conclusions can
be reached.

This paper is organized as follows. In the next section we describe in
detail the relevant group theory associated with the global symmetry group
$O(m)^n\rtimes S_n$ and derive the associated crossing equation. In section
\ref{secMNAnis} we briefly mention results of the $\veps$ expansion for
theories with $O(m)^n\rtimes S_n$ symmetry and some of the physical systems
such theories are expected to describe at criticality. In section
\ref{secTetraSym} we turn to the group theory of the global symmetry group
$D_4{\!}^n\rtimes S_n$ and we derive the crossing equation for this case.
In section \ref{secTetraAnis} we mention some aspects of the application of
the $\veps$ expansion to theories with $D_4{\!}^n\rtimes S_n$ symmetry.
Finally, we present our numerical results in section \ref{secNumRes} and
conclude in section \ref{secConc}.

\newsec{MN symmetry}[MNsym]
Let us recall some basic facts about semidirect products. To have a
well-defined semidirect product, $G=N\rtimes H$, with $N, H$ subgroups of
$G$ with $H$ proper and $N$ normal, i.e.\ $H\subset G$ and $N\lhd G$, we
need to specify the action of $H$ on the group of automorphisms of $N$.
This action is defined by a map $f:H\to\text{Aut}(N)$, $f: h\mapsto
f(h)=f_h$. The action of $f_h$ on $N$ is given by conjugation, $f_h:N\to
N$, $f_h:n\mapsto hnh^{-1}$.  (By definition $hnh^{-1}\in N$ since $N\lhd
G$.) With this definition, $f$ is a homomorphism, i.e.\
$f_{h_1}f_{h_2}=f_{h_1h_2}$.  One can show that, up to isomorphisms, $N,H$
and $f$ uniquely determine $G$. The multiplication of two elements $(n,h)$
and $(n',h')$ of $G$ is given by
\eqn{(n,h)(n',h')=(nf_h(n'),hh')\,,}[]
the identity element is $(e_N,e_H)$, with $e_N$ the identity element of $N$
and $e_H$ that of $H$, and the inverse of $(n,h)$ is given by
\eqn{(n,h)^{-1}=(f_{h^{-1}}(n^{-1}),h^{-1})\,.}[]
Note that a direct product is a special case of a semidirect product where
$f$ is the trivial homomorphism, i.e.\ the homomorphism that sends every
element of $H$ to the identity automorphism of $N$.

In this work we analyze CFTs with global symmetry of the form $K^n\rtimes
S_n$, where $K^n$ denotes the direct product of $n$ groups $K$ and $S_n$
the permutation group of $n$ elements. In this case the action of the
homomorphism $f:S_n\to \text{Aut}(K^n)$ is to permute the $K$'s in $K^n$,
i.e.\ $f_{\sigma}:(k_1,\ldots,k_n)\mapsto
(k_{\sigma(1)},\ldots,k_{\sigma(n)})$, with $\sigma$ an element of $S_n$
and $k_i$, $i=1,\ldots,n$, an element of the $i$-th $K$ in $K^n$.\foot{This
type of semidirect product is an example of a wreath product, for which the
standard notation is $K\wr S_n$.}

The first example we analyze is that of the $\MN_{m,n}$ CFT. By this we
refer to the CFT with global symmetry $\MN_{m,n}=O(m)^n\rtimes S_n$.  The
vector representation is furnished by the operator $\phi_i$,
$i=1,\ldots,mn$. The crucial group-theory problem is of course to decompose
$\langle\phi_i(x_1)\phi_j(x_2)\phi_k(x_3)\phi_l(x_4)\rangle$ into invariant
subspaces in order to derive the set of crossing equations that constitutes
the starting point for our numerical analysis. Invariant tensors help us in
this task. As far as the OPE is concerned we have
\eqn{\phi_i\times\phi_j\sim\delta_{ij}S+X_{(ij)}+Y_{(ij)}+Z_{(ij)}
+A_{[ij]}+B_{[ij]}\,,}[]
where $S$ is the singlet, $X,Y,Z$ are traceless-symmetric and $A,B$
antisymmetric.

If one thinks of the symmetry breaking $O(mn)\to\MN_{m,n}$, then the
irreducible representations (irreps) $X,Y,Z$ stem from the
traceless-symmetric irrep of $O(mn)$, while $A,B$ stem from the
antisymmetric irrep of $O(mn)$. The way to figure out the explicit way the
$O(mn)$ representations decompose under the action of the $\MN_{m,n}$ group
is by constructing the appropriate projectors. The first step to doing that
is to construct the invariant tensors of the group under study. This way of
thinking, in terms of invariant theory, was recently applied to the $\veps$
expansion in~\cite{Osborn:2017ucf}, and it turns out to be very useful when
thinking about the problem from the bootstrap point of view.

\subsec{Invariant tensors and projectors}
For the $\MN_{m,n}$ CFT there are two four-index primitive invariant
tensors~\cite{Osborn:2017ucf}. They can be defined as follows:
\twoseqn{&\gamma_{ijkl}\lsp\phi_i\phi_j\phi_k\phi_l=
(\phi_1^2+\cdots+\phi_m^2)^2
+(\phi_{m+1}^2+\cdots+\phi_{2m}^2)^2
+\cdots+(\phi_{m(n-1)+1}^2+\cdots+\phi_{mn}^2)^2\,,}[gamzetI]{
&\!\omega_{ijkl}\lsp\phi_i\phi_j\phi'_k\phi'_l=
(\phi_1\phi'_2-\phi_2\phi'_1)^2+(\phi_3\phi'_4-\phi_4\phi'_3)^2
+\cdots+(\phi_{mn-1}\phi'_{mn}-\phi_{mn}\phi'_{mn-1})^2\,.
}[gamzetII][gamzet]
The tensor $\gamma$ is fully symmetric, while the tensor $\omega$ satisfies
\eqn{\omega_{ijkl}=\omega_{jikl}\,,\qquad
\omega_{ijkl}=\omega_{klij}\,,\qquad
\omega_{ijkl}+\omega_{ikjl}+\omega_{iljk}=0\,.}[zetaSym]
A non-primitive invariant tensor with four indices is defined by
\eqn{\xi_{ijkl}\lsp\phi_i\phi_j\phi_k\phi_l=(\phi_1^2+\phi_2^2+\cdots
+\phi_{mn}^2)^2\,,}[invXi]
which respects $O(mn)$ symmetry.  One can verify that (repeated indices are
always assumed to be summed over their allowed values)
\eqn{\gamma_{iijk}=\tfrac13(m+2)\lsp\delta_{jk}\,,\qquad
\omega_{iijk}=(m-1)\lsp\delta_{jk}\,,}[trI]
and
\eqna{\gamma_{ijmn}\gamma_{klmn}&=\tfrac{1}{9}(m+8)\lsp\gamma_{ijkl}
+\tfrac{2}{27}(m+2)\lsp\omega_{ijkl}\,,\\
\gamma_{ijmn}\omega_{klmn}&=\tfrac13(m-1)\lsp\gamma_{ijkl}
+\tfrac{2}{9}(m+2)\lsp\omega_{ijkl}\,,\\
\omega_{ijmn}\omega_{klmn}&=(m-1)\lsp\gamma_{ijkl}
+\tfrac13(2m-5)\lsp\omega_{ijkl}\,,\\
\omega_{imjn}\omega_{kmln}&=\tfrac14(m-1)\lsp\gamma_{ijkl}
+\tfrac16(m+2)\lsp\omega_{ijkl}+\tfrac32\lsp\omega_{ikjl}\,.}[relI]
With the help of \trI and \relI it can be shown that the tensors
\begin{subequations}\label{invITen}
\begin{equation}\label{invITenI}
  P^S_{ijkl}=\tfrac{1}{mn}\lsp\delta_{ij}\delta_{kl}\,,
\end{equation}\vspace{-21pt}
\begin{equation}\label{invITenII}
  P^X_{ijkl}=\tfrac{1}{m}\lsp\gamma_{ijkl} + \tfrac{2}{3m}\lsp
  \omega_{ijkl}-\tfrac{1}{mn}\lsp\delta_{ij}\delta_{kl}\,,
\end{equation}
\begin{equation}\label{invITenIII}
  P^Y_{ijkl}=(1-\tfrac{1}{m})\lsp\gamma_{ijkl}
  -\tfrac13(1+\tfrac{2}{m})\lsp\omega_{ijkl}\,,
\end{equation}
\begin{equation}\label{invITenIV}
  P^Z_{ijkl}=-\gamma_{ijkl}+\tfrac13\lsp\omega_{ijkl}
  +\tfrac12(\delta_{ik}\delta_{jl} +\delta_{il}\delta_{jk})\,,
\end{equation}
\begin{equation}\label{invITenV}
  P^A_{ijkl}=\tfrac13(\omega_{ijkl}+2\lsp\omega_{ikjl})\,,
\end{equation}
\begin{equation}\label{invITenVI}
  P^B_{ijkl}=-\tfrac13(\omega_{ijkl}+2\lsp\omega_{ikjl})+\tfrac12
  (\delta_{ik}\delta_{jl}-\delta_{il}\delta_{jk})\,,
\end{equation}
\end{subequations}
satisfy
\eqn{P_{ijmn}^{I}P_{mnkl}^{J}=P_{ijkl}^{I}\lsp\delta^{IJ}\,,\qquad
\sum_{I}P^{I}_{ijkl}=\delta_{ik}\delta_{jl}\,,\qquad
P_{ijkl}^{I}\lsp\delta_{ik}\delta_{jl}=d_r^{I}\,,}[orthI]
where $d_r^{I}$ is the dimension of the representation indexed by $I$, with
\eqn{\{d_r^S, d_r^X, d_r^Y, d_r^Z, d_r^A, d_r^B\}
=\{1,n-1,\tfrac12(m-1)(m+2)n,\tfrac12m^2n(n-1),
\tfrac12mn(m-1),\tfrac12m^2n(n-1)\}\,.}[]
The dimensions $d_r^X,d_r^Y,d_r^Z$ are as expected from the results of
\cite[Eq.~(5.98)]{Osborn:2017ucf}.

Knowledge of the projectors (\ref{invITen}a--f) allows the derivation of
the corresponding crossing equation in the usual way. The four-point
function can be expressed in a conformal block decomposition in the
$12\rightarrow34$ channel as
\eqn{\langle\phi_i(x_1)\phi_j(x_2)\phi_k(x_3)\phi_l(x_4)\rangle=
\frac{1}{(x_{12}^2 x_{34}^2)^{\Delta_\phi}}
\sum_{I}\sum_{\cO_I} \lambda_{\cO_I}^2
P^I_{ijkl}\lsp g_{\Delta_I,\ell_I}(u,v)\,,}[]
where the sum over $I$ runs over the representations $S,X,Y,Z,A,B$,
$x_{ij}=x_i-x_j$, $\lambda_{\cO_I}^2$ are squared OPE coefficients and
$g_{\Delta_I,\ell_I}(u,v)$ are conformal blocks\foot{We define conformal
blocks using the conventions of~\cite{Behan:2016dtz}.} that are functions
of the conformally-invariant cross ratios
\eqn{u=\frac{x_{12}^2x_{34}^2}{x_{13}^2x_{24}^2}\,,\qquad
v=\frac{x_{14}^2x_{23}^2}{x_{13}^2x_{24}^2}\,.}[]
The crossing equation can now be derived. With
\eqn{F_{\Delta,\lsp\ell}^{\pm}(u,v)=v^{\Delta_\phi}g_{\Delta,\lsp\ell}(u,v)
\pm u^{\Delta_\phi}g_{\Delta,\lsp\ell}(v,u)\,,}[Fpmdef]
we find\foot{In \eqref{crEqMN} we omit, for brevity, to label the
$F_{\Delta,\ell}$'s and $\lambda_\cO^2$'s with the appropriate index $I$.
The appropriate labeling, however, is obvious from the overall sum in each
term.}

\eqna{&\sum_{S^+}\lambda_\cO^2\begin{pmatrix}
  F^-_{\Delta,\lsp\ell}\\
  0\\
  0\\
  0\\
  F^+_{\Delta,\lsp\ell}\\
  0
\end{pmatrix}+
\sum_{X^+}\lambda_\cO^2\begin{pmatrix}
  -F^-_{\Delta,\lsp\ell}\\
  F^-_{\Delta,\lsp\ell}\\
  0\\
  0\\
  -F^+_{\Delta,\lsp\ell}\\
  F^+_{\Delta,\lsp\ell}
\end{pmatrix}+
\sum_{Y^+}\lambda_\cO^2\begin{pmatrix}
  0\\
  \tfrac{m-1}{n}\lsp F^-_{\Delta,\lsp\ell}\\
  F^-_{\Delta,\lsp\ell}\\
  0\\
  0\\
  -\tfrac{m+2}{2n}\lsp F^+_{\Delta,\lsp\ell}
\end{pmatrix}
+\sum_{Z^+}\lambda_\cO^2\begin{pmatrix}
  0\\
  0\\
  0\\
  F^-_{\Delta,\lsp\ell}\\
  -\tfrac12\lsp F^+_{\Delta,\lsp\ell}\\
  \tfrac{1}{2n}\lsp F^+_{\Delta,\lsp\ell}
\end{pmatrix}\\
&\hspace{6cm}+
\sum_{A^-}\lambda_\cO^2\begin{pmatrix}
  0\\
  0\\
  \tfrac{1}{m}\lsp F^-_{\Delta,\lsp\ell}\\
  0\\
  0\\
  \tfrac{1}{2n}\lsp F^+_{\Delta,\lsp\ell}
\end{pmatrix}+
\sum_{B^-}\lambda_\cO^2\begin{pmatrix}
  -F^-_{\Delta,\lsp\ell}\\
  \tfrac{1}{n}\lsp F^-_{\Delta,\lsp\ell}\\
  0\\
  F^-_{\Delta,\lsp\ell}\\
  \tfrac12\lsp F^+_{\Delta,\lsp\ell}\\
  -\tfrac{1}{2n}\lsp F^+_{\Delta,\lsp\ell}
\end{pmatrix}=
\begin{pmatrix}
  0\\
  0\\
  0\\
  0\\
  0\\
  0
\end{pmatrix}.}[crEqMN]
The signs that appear as superscripts in the various irrep symbols indicate
the spins of the operators we sum over in the corresponding term: even when
positive and odd when negative.

\newsec{MN anisotropy}[secMNAnis]
The $\MN_{m,n}$ fixed points were first studied in~\cite{Shpot, Shpot2,
Mudrov, Mudrov:2001yr, Mudrov2} and more recently in \cite{Osborn:2017ucf,
Rychkov:2018vya}. The relevant Lagrangian is\foot{Compared to couplings
$\lambda,g$ of \cite[Sec.~5.2.2]{Osborn:2017ucf} we have
$\lambda^{\text{here}}=\lambda^{\text{there}}-\frac{m+2}{3(mn+2)}g^{\text{there}}$
and $g^{\text{here}}=g^{\text{there}}$.}
\eqn{\mathscr{L}=\tfrac12\lsp\partial_\mu\phi_i\lsp\partial^\mu\phi_i
  +\tfrac18\lsp(\lambda\lsp\xi_{ijkl} +\tfrac13\lsp g\lsp\gamma_{ijkl})
\lsp\phi_i\phi_j\phi_k\phi_l\,.}[LagMN]
In the $\veps$ expansion below $d=4$ \LagMN has four inequivalent fixed
points. They are
\begin{enumerate}
  \item Gaussian ($\lambda=g=0$),
  \item $O(mn)$ ($\lambda>0,g=0$)
  \item $n$ decoupled $O(m)$ models ($\lambda=0$, $g>0$),
  \item $n$ coupled $O(m)$ models with symmetry
    $\MN_{m,n}=O(m)^n\rtimes S_n$ ($\lambda>0$, $g\neq0$).\foot{Although
    the theory of $n$ decoupled $O(m)$ models in item 3 on the list also
    has symmetry $\MN_{m,n}$, we will never characterize it that way; we
    will reserve that characterization for the fully-interacting case in
    4.}
\end{enumerate}

These fixed points may be physically relevant for $m=2$ and $n=2,3$. In the
$\veps$ expansion, the $\MN_{2,2}$ CFT is equivalent to a theory with
$O(2)^2/\mathbb{Z}_2$ symmetry~\cite{Pelissetto:2000ek, Osborn:2017ucf,
Rychkov:2018vya}. Lagrangians with $O(2)\times O(n)/\mathbb{Z}_2$ symmetry
have fixed points with collinear (also referred to as sinusoidal) or
noncollinear (also referred to as chiral) order depending on
$n$~\cite{Kawamura_1998, Pelissetto:2000ek}. The chiral (resp.\ sinusoidal)
region is defined by the requirement $g<0$ (resp.\ $g>0$), which
corresponds to $v>0$ (resp. $v<0$) in the notation of~\cite{Kawamura_1998}.
The $\MN_{2,2}$ fixed point may have applications to XY stacked triangular
antiferromagnets and paramagnetic-helimagnetic transitions in the
rare-earth metals Ho (holmium), Dy (dysprosium) and Tb (terbium). This
requires $g<0$, which is in tension with extrapolations derived from the
$\veps$ expansion. As a result, the relevance of the $\MN_{2,2}$ fixed
point found with the $\veps$ expansion for these transitions is not widely
accepted~\cite[Sec.~11.5]{Pelissetto:2000ek}. This fixed point has also
been argued to describe the structural phase transition of NbO$_2$ (niobium
dioxide).

The $\MN_{2,3}$ fixed point could be relevant for the antiferromagnetic
phase transitions in K$_2$IrCl$_6$ (potassium hexachloroiridate), TbD$_2$
(terbium dideuteride) and Nd (neodymium)~\cite{Mukamel0, Mukamel1,
Mukamel2, Mukamel3, Bak:1976zz}.

The stability of the $\MN_{m,n}$ fixed point for $m=2$ and $n=2,3$ has been
supported by higher-loop $\varepsilon$ expansion calculations~\cite{Mudrov,
Mudrov:2001yr, Mudrov2}.  However, there exist higher-loop calculations
based on the fixed-dimension expansion---see \cite{Pelissetto:2000ek} and
references therein---indicating that the stable fixed point is actually
that of $n$ decoupled $O(2)$ models. In $d=3$, the theory of $n$ decoupled
$O(2)$ models is nonperturbatively stable, due to the fact that in the
decoupled $O(2)$ model the potentially relevant perturbation has dimension
twice that of the leading scalar singlet of the $O(2)$ model, which in turn
has dimension slightly above 1.5~\cite{Kos:2016ysd}. As mentioned in the
introduction, our numerical results indicate that the $\MN_{2,2}$ and
$\MN_{2,3}$ theories of the $\veps$ expansion are both stable (assuming
that our corresponding kinks correspond to the $\veps$ expansion fixed
points). This suggests either that there are two stable fixed points in
$d=3$, in direct contradiction with intuition derived from the $\veps$
expansion~\cite{Michel:1983in, Rychkov:2018vya}, or that our spectrum
analysis misses the slightly relevant operator expected from the $\veps$
expansion. It is also possible that our numerical results for $\MN_{2,2}$
and $\MN_{2,3}$ theories do not actually pertain to the fixed points of the
$\veps$ expansion---in that case our results would signify the discovery of
new CFTs. Indeed, application of our $\MN_{2,2}$ results to physical
systems requires $g<0$ which may be in tension with the $\veps$ expansion
as described above. We are unable to conclusively resolve these issues in
this work.

\newsec{Tetragonal symmetry}[secTetraSym]
The tetragonal CFT~\cite{Pelissetto:2000ek, Osborn:2017ucf} has global
symmetry $R_n=D_4{\!}^n\rtimes S_n$, where $D_4$ is the eight-element
dihedral group. For $n=0$ $R_0=\{e\}$, where $e$ is the identity element,
and for $n=1$ $R_1=D_4$.  The order of $R_n$ is $\ord(R_n)=8^n n!$.  Note
that $R_n$ is a subgroup of the hypercubic group
$C_N=\mathbb{Z}_2{\!}^N\rtimes S_N$, $N=2n$, whose order is $\ord(C_N)=2^N
N!$. It is easy to see that $\ord(C_N)/\ord(R_n)=(2n-1)!!$, which is an
integer for any integer $n\ge0$.

The number of irreps of the group $R_n$ for $n=0,1,2,3,4,5,\ldots$ is $1,
5, 20, 65, 190, 506,\ldots,$ respectively.\foot{These numbers have been
obtained with the use of the freely available software
\href{https://www.gap-system.org}{\texttt{GAP}}~\cite{GAP4}.} Among the
irreps of $R_n$ one always finds a $2n$-dimensional one; we will refer to
this as the vector representation $\phi_i, i=1,\ldots,2n$.

In this work we analyze bootstrap constraints on the four-point function of
the vector operator $\phi_i$. A standard construction of the character
table shows that the group $R_2$ has eight one-dimensional, six
two-dimensional and six four-dimensional irreps.\foot{Character tables for
a wide range of finite groups can be easily generated using
\href{https://www.gap-system.org}{\texttt{GAP}}~\cite{GAP4}.} In this case
we may write\foot{Of course these $S,X,Y,Z,A,B$ have nothing to do with the
ones of section \MNsym.}
\eqn{\overset{\,4}{\phi}_i\times\overset{\,4}{\phi}_j\sim
\delta_{ij}\overset{\lsp 1}{S}+\overset{2}{W}_{\!(ij)}
+\overset{\,1}{X}_{\!(ij)}+\overset{2}{Y}_{\!(ij)}
+\overset{\,4}{Z}_{(ij)}+\overset{\;2}{A}_{[ij]}
+\overset{\,4}{B}_{[ij]}\,.}[OPEI]
$S$ is the singlet. The dimensions of the various irreps are given by the
number over their symbol.  $W,X,Y,Z$ are two-index symmetric and traceless,
while $A,B$ are two-index antisymmetric.

\subsec{Invariant tensors}
In the tetragonal case there are three primitive invariant tensors with
four indices, defined by
\threeseqn{\delta_{ijkl}\lsp\phi_i\phi_j\phi_k\phi_l&=\phi_1^4+\phi_2^4
+\cdots+\phi_{2n}^4\,,}[primInvI]{\zeta_{ijkl}\lsp\phi_i\phi_j\phi_k\phi_l
&=2\lsp(\phi_1^2\phi_2^2+\phi_3^2\phi_4^2+\cdots+\phi_{2n-1}^2\phi_{2n}^2)
\,,}[primInvII]{\omega_{ijkl}\lsp\phi_i\phi_j\phi'_k\phi'_l&=(\phi_1\phi'_2
-\phi_2\phi'_1)^2+(\phi_3\phi'_4-\phi_4\phi'_3)^2+\cdots
+(\phi_{2n-1}\phi'_{2n}-\phi_{2n}\phi'_{2n-1})^2\,.}[primInvIII][primInv]
The tensors $\delta, \zeta$ are fully symmetric, while the tensor $\omega$
is the same as that in \gamzetII for $m=2$.  It can be verified that these
satisfy
\eqn{\delta_{iijk}=3\lsp\zeta_{iijk}=\omega_{iijk}
=\delta_{jk}\,,}[trRel]
and
\eqna{\delta_{ijmn}\delta_{klmn}&=\delta_{ijkl}\,,\\
\delta_{ijmn}\zeta_{klmn}&=\tfrac13\lsp\zeta_{ijkl}
+\tfrac29\lsp\omega_{ijkl}\,,\\
\delta_{ijmn}\omega_{klmn}&=\zeta_{ijkl}+\tfrac23\lsp\omega_{ijkl}\,,\\
\zeta_{ijmn}\zeta_{klmn}&=\tfrac19\lsp\delta_{ijkl}
+\tfrac49\lsp\zeta_{ijkl}-\tfrac{4}{27}\lsp\omega_{ijkl}\,,\\
\zeta_{ijmn}\omega_{klmn}&=\tfrac13\lsp\delta_{ijkl}
-\tfrac23\lsp\zeta_{ijkl}+\tfrac29\lsp\omega_{ijkl}\,,\\
\omega_{ijmn}\omega_{klmn}&=\delta_{ijkl}+\zeta_{ijkl}
-\tfrac13\lsp\omega_{ijkl}\,,\\
\omega_{imjn}\omega_{kmln}&=\tfrac14\lsp\delta_{ijkl}
+\tfrac14\lsp\zeta_{ijkl}+\tfrac23\lsp\omega_{ijkl}
+\tfrac32\lsp\omega_{ikjl}\,.}[rels]
These relations are valid for any $n\ge2$.

To verify that there are only three invariant polynomials of $R_n$ made out
of the components of the vector $\phi_i$, we have computed the Molien
series for $n=2,3,4$.\foot{For $n=3,4$ the computation of the Molien series
was performed with
\href{https://www.gap-system.org}{\texttt{GAP}}~\cite{GAP4}.} To do this,
we think of $R_n$ as represented by $2n\times 2n$ matrices acting on the
$2n$-component vector $\phi_i^T$. Using those matrices, which represent the
group elements $g_i\in G$ as $\rho(g_i)$, $i=1,\ldots,\ord(G)$, we can then
explicitly compute the Molien series. The Molien formula is
\eqn{M(t)=\frac{1}{\ord(G)}\sum_{i=1}^{\ord(G)}\frac{1}{\det(\mathds{1}-t
\lsp\rho(g_i))}\,,}[MolFor]
where $\mathds{1}$ is the identity matrix of appropriate size. It is
obvious that the summands in \MolFor only depend on the conjugacy class, so
the sum can be taken to be over conjugacy classes with the appropriate
weights. For $n=2,3,4$ \MolFor gives, respectively,
\eqna{M_2(t)&=\frac{t^4-t^2+1}{(t^4+1)^2(t^2+1)^2(t^2-1)^4}\,,\\
M_3(t)&=\frac{t^{16}-t^{14}+t^{12}+t^8+t^4-t^2+1}
{(t^4+t^2+1)^2(t^4-t^2+1)^2(t^4+1)(t^2+1)^3(t^2-1)^6}\,,\\
M_4(t)&=\frac{(t^{20}-t^{18}+t^{14}+t^{12}-t^{10}+t^{8}+t^{6}-t^2+1)
(t^8-t^6+t^4-t^2+1)}{(t^8+1)(t^4-t^2+1)(t^4+1)^2(t^2+t+1)^2
(t^2-t+1)^2(t^2+1)^4(t^2-1)^8}\,,}[]
whose series expansions are
\eqna{M_2(t)&=1+t^2+3\lsp t^4+4\lsp t^6+8\lsp
t^8+\text{O}(t^{10})\,,\\
M_3(t)&=1+t^2+3\lsp t^4+5\lsp t^6+10\lsp t^8+\text{O}(t^{10})\,,\\
M_4(t)&=1+t^2+3\lsp t^4+5\lsp t^6+11\lsp t^8+\text{O}(t^{10})\,.}[]
Thus, we see that we have one quadratic and three quartic invariants. The
latter are generated by $\delta_{ijkl}$, $\zeta_{ijkl}$ and $\xi_{ijkl}$,
and their form is given in (\ref{primInv}a,b) and \invXi with $m=2$.  The
unique quadratic invariant is obviously generated by $\delta_{ij}$ and it
is given by $\phi^2=\phi_1^2+\phi_2^2+\cdots+\phi_{2n}^2$.

\subsec{Projectors and crossing equation}
If we now define
\begin{subequations}\label{invTen}
\begin{equation}\label{invTenI}
  \hP^S_{ijkl}=\tfrac{1}{2n}\delta_{ij}\delta_{kl}\,,
\end{equation}
\begin{equation}\label{invTenII}
  \hP^W_{ijkl}=\tfrac12(\delta_{ijkl}-\zeta_{ijkl})
  -\tfrac13\lsp\omega_{ijkl}\,,
\end{equation}
\begin{equation}\label{invTenIII}
  \hP^X_{ijkl}=\tfrac12(\delta_{ijkl}+\zeta_{ijkl})
  +\tfrac13\lsp\omega_{ijkl}-\tfrac{1}{2n}\delta_{ij}\delta_{kl}\,,
\end{equation}
\begin{equation}\label{invTenIV}
  \hP^Y_{ijkl}=\zeta_{ijkl}-\tfrac13\lsp\omega_{ijkl}\,,
\end{equation}
\begin{equation}\label{invTenV}
  \hP^Z_{ijkl}=-\delta_{ijkl}-\zeta_{ijkl}+\tfrac13\lsp\omega_{ijkl}
  +\tfrac12(\delta_{ik}\delta_{jl} +\delta_{il}\delta_{jk})\,,
\end{equation}
\begin{equation}\label{invTenVI}
  \hP^A_{ijkl}=\tfrac13(\omega_{ijkl}+2\lsp\omega_{ikjl})\,,
\end{equation}
\begin{equation}\label{invTenVII}
  \hP^B_{ijkl}=-\tfrac13(\omega_{ijkl}+2\lsp\omega_{ikjl})+\tfrac12
  (\delta_{ik}\delta_{jl}-\delta_{il}\delta_{jk})\,,
\end{equation}
\end{subequations}
we may verify, using \trRel and \rels, the projector relations
\eqn{\hP_{ijmn}^{I}\hP_{mnkl}^{J}=\hP_{ijkl}^{I}\lsp\delta^{IJ}\,,\qquad
\sum_{I}\hP^{I}_{ijkl}=\delta_{ik}\delta_{jl}\,,\qquad
\hP_{ijkl}^{I}\lsp\delta_{ik}\delta_{jl}=\hd_r^{I}\,,}[orth]
where $\hd_r^{I}$ is the dimension of the representation indexed by $I$,
with
\eqn{\{\hd_r^S, \hd_r^W, \hd_r^X, \hd_r^Y, \hd_r^Z, \hd_r^A, \hd_r^B\}
=\{1,n,n-1,n,2n(n-1),n,2n(n-1)\}\,.}[]
The generalization of \OPEI, valid for any $n\ge2$, is
\eqn{\overset{2n}{\phi_i}\times\overset{2n}{\phi_j}\sim
\delta_{ij}\overset{\;1}{S}+\overset{\,n}{W}_{\!(ij)}
+\overset{\,n-1}{X}_{\!\!\!(ij)}+\overset{\,n}{Y\!}_{(ij)}
+\hspace{-6pt}\overset{2n(n-1)}{Z}_{\!\!\!\!\!\!\!\!\!(ij)}
+\overset{\,\,n}{A}_{[ij]}
+\hspace{-6pt}\overset{2n(n-1)}{B}_{\!\!\!\!\!\!\!\![ij]}\,.}[OPEII]

The projectors (\ref{invTen}a--g) allow us to express the four-point function
of interest in a conformal block decomposition in the $12\rightarrow34$
channel:
\eqn{\langle\phi_i(x_1)\phi_j(x_2)\phi_k(x_3)\phi_l(x_4)\rangle=
\frac{1}{(x_{12}^2 x_{34}^2)^{\Delta_\phi}}
\sum_{I}\sum_{\cO_I} \lambda_{\cO_I}^2
\hP^I_{ijkl}\lsp g_{\Delta_I,\ell_I}(u,v)\,,}[]
where the sum over $I$ runs over the representations $S,W,X,Y,Z,A,B$. For
the crossing equation we find\foot{In \eqref{crEqTetra} we omit, for
brevity, to label the $F_{\Delta,\ell}$'s and $\lambda_\cO^2$'s with the
appropriate index $I$.  The appropriate labeling, however, is obvious from
the overall sum in each term.}
\eqna{&\sum_{S^+}\lambda_\cO^2\begin{pmatrix}
  F^-_{\Delta,\lsp\ell}\\
  0\\
  0\\
  0\\
  0\\
  F^+_{\Delta,\lsp\ell}\\
  F^+_{\Delta,\lsp\ell}
\end{pmatrix}+
\sum_{W^+}\lambda_\cO^2\begin{pmatrix}
  0\\
  F^-_{\Delta,\lsp\ell}\\
  0\\
  0\\
  0\\
  -F^+_{\Delta,\lsp\ell}\\
  0
\end{pmatrix}+
\sum_{X^+}\lambda_\cO^2\begin{pmatrix}
  -\tfrac{1}{n}F^-_{\Delta,\lsp\ell}\\
  F^-_{\Delta,\lsp\ell}\\
  F^-_{\Delta,\lsp\ell}\\
  -F^-_{\Delta,\lsp\ell}\\
  0\\
  (1-\tfrac{1}{n})F^+_{\Delta,\lsp\ell}\\
  -\tfrac{1}{n}F^+_{\Delta,\lsp\ell}
\end{pmatrix}+
\sum_{Y^+}\lambda_\cO^2\begin{pmatrix}
  0\\
  0\\
  F^-_{\Delta,\lsp\ell}\\
  0\\
  0\\
  -F^+_{\Delta,\lsp\ell}\\
  0
\end{pmatrix}\\
&\hspace{4cm}+
\sum_{Z^+}\lambda_\cO^2\begin{pmatrix}
  2\llsp F^-_{\Delta,\lsp\ell}\\
  -2\llsp F^-_{\Delta,\lsp\ell}\\
  -2\llsp F^-_{\Delta,\lsp\ell}\\
  2\llsp F^-_{\Delta,\lsp\ell}\\
  F^-_{\Delta,\lsp\ell}\\
  0\\
  -F^+_{\Delta,\lsp\ell}
\end{pmatrix}+
\sum_{A^-}\lambda_\cO^2\begin{pmatrix}
  0\\
  0\\
  0\\
  F^-_{\Delta,\lsp\ell}\\
  0\\
  F^+_{\Delta,\lsp\ell}\\
  0
\end{pmatrix}+
\sum_{B^-}\lambda_\cO^2\begin{pmatrix}
  0\\
  0\\
  0\\
  0\\
  F^-_{\Delta,\lsp\ell}\\
  0\\
  F^+_{\Delta,\lsp\ell}
\end{pmatrix}=
\begin{pmatrix}
  0\\
  0\\
  0\\
  0\\
  0\\
  0\\
  0
\end{pmatrix}.}[crEqTetra]

Let us make a comment about \crEqTetra. We observe that we obtain the same
crossing equation if we exchange the second and third line in all vectors
and at the same time relabel $W^+\leftrightarrow Y^+$. This implies, for
example, that operator dimension bounds on the leading scalar $W$ operator
and the leading scalar $Y$ operator will be identical.  Furthermore, if we
work out the spectrum on the $W$- and the $Y$-bound, then all operators in
the solution will have the same dimensions in both cases (except for the
relabeling $W^+\leftrightarrow Y^+$). The reason for this is that there
exists a transformation of $\phi_i$ that permutes the projectors $\hP^W$
and $\hP^X$.\foot{This was suggested to us by Hugh Osborn.} Indeed, if
\eqn{\phi_i\to\tfrac{1}{\sqrt{2}}(\phi_i+\phi_{i+1})\,,\quad i\text{
  odd}\quad\text{and}\quad
  \phi_i\to\tfrac{1}{\sqrt{2}}(\phi_{i-1}-\phi_i)\,,\quad i\text{
even}\,,}[sym]
then
\eqn{\delta_{ij}\to\delta_{ij}\,,\quad
\delta_{ijkl}\to\tfrac12(\delta_{ijkl}+3\lsp\zeta_{ijkl})\,,
\quad\zeta_{ijkl}\to\tfrac12(\delta_{ijkl}-\zeta_{ijkl})
\quad\text{and}\quad \omega_{ijkl}\to\omega_{ijkl}\,.}[trans]
Under \trans we obviously have $\hP^W\leftrightarrow\hP^Y$. Let us remark
here that something similar happens in the $N=2$ cubic theory studied
in~\cite[Sec.~6]{Stergiou:2018gjj}, again due to the transformation \sym
that exchanges two projectors.\foot{In the $N=2$ cubic case, which
corresponds to $n=1$ here in which case the $\zeta$ tensor does not exist,
we can show that $\delta_{ij}\to\delta_{ij}$ and
$\delta_{ijkl}\to-\delta_{ijkl}+\tfrac12(\delta_{ij}\delta_{kl}
+\delta_{ik}\delta_{kl}+\delta_{il}\delta_{jk}\llnsp)$.}

With the crossing equation \crEqTetra we can now commence our numerical
bootstrap explorations. Before that, however, let us first summarize
results of the $\veps$ expansion for theories with tetragonal anisotropy.

\newsec{Tetragonal anisotropy}[secTetraAnis]
Theories with tetragonal anisotropy were first studied with the
$\varepsilon$ expansion a long time ago in \cite{Mukamel0, Mukamel1,
Mukamel2, Mukamel3, Bak:1976zz} and later \cite{Mudrov3}, and they were
revisited recently in~\cite{Osborn:2017ucf, Rychkov:2018vya}. A standard
review is~\cite[Sec.~11.6]{Pelissetto:2000ek}. The Lagrangian one starts
with is\foot{Compared to couplings $\lambda,g_1,g_2$ of
\cite[Sec.~7]{Osborn:2017ucf} we have
$\lambda^{\text{here}}=\lambda^{\text{there}}-\frac{2}{3(n+1)}
g^{\text{there}}$, $g_1^{\text{here}}=g_1^{\text{there}}$ and
$g_2^{\text{here}}=g_2^{\text{there}}$.}
\eqn{\mathscr{L}=\tfrac12\lsp\partial_\mu\phi_i\lsp\partial^\mu\phi_i
+\tfrac18\lsp(\lambda\lsp\xi_{ijkl} +\tfrac13\lsp g_1\lsp\delta_{ijkl}
+\tfrac13\lsp g_2\lsp\zeta_{ijkl})\lsp\phi_i\phi_j\phi_k\phi_l
\,.}[Lag]
For $g_1=g_2=g$ this reduces to \LagMN. The theory \Lag in
$d=4-\varepsilon$ has six inequivalent fixed points.  They are\foot{Fixed
points physically-equivalent to those in items 2 and 5 on the list are also
found in other positions in coupling space, related to the ones given in
the list by the field redefinition in \sym~\cite{Pelissetto:2000ek,
Osborn:2017ucf}.}
\begin{enumerate}
  \item Gaussian ($\lambda=g_1=g_2=0$),
  \item $2n$ decoupled Ising models ($\lambda=g_2=0$, $g_1>0$),
  \item $n$ decoupled $O(2)$ models ($\lambda=0$, $g_1=g_2>0$),
  \item $O(2n)$ ($\lambda>0$, $g_1=g_2=0$),
  \item Hypercubic with symmetry $C_{2n}=\mathbb{Z}_2{\!}^{\lsp 2n}\rtimes
    S_{2n}$ ($\lambda>0$, $g_1>0$, $g_2=0$),\foot{The theory of $2n$
    decoupled Ising models in item 2 on the list has symmetry $C_{2n}$ as
    well.  However, we reserve the $C_{2n}$ characterization for the theory
    in 5.}
  \item $n$ coupled $O(2)$ models with symmetry
    $\MN_{2,n}=O(2)^n\rtimes S_n$ ($\lambda>0$, $g_1=g_2\ne0$).\foot{The
      theory of $n$ decoupled $O(2)$ models in item 3 on the list has
      symmetry $\MN_{2,n}$ as well. However, we reserve the $\MN_{2,n}$
      characterization for the theory in item 6.}
\end{enumerate}

Note that in the $\varepsilon$ expansion there is no $R_n$ symmetric fixed
point. According to the $\varepsilon$ expansion the stable fixed point is
the $\MN_{2,n}$ symmetric one we discussed in section~\secMNAnis.

\newsec{Numerical results}[secNumRes]
The numerical results in this paper have been obtained with the use of
\href{https://github.com/cbehan/pycftboot}{\texttt{PyCFTBoot}}~\cite{Behan:2016dtz}
and
\href{https://github.com/davidsd/sdpb}{\texttt{SDPB}}~\cite{Simmons-Duffin:2015qma}.
We use $\texttt{nmax}=9$, $\texttt{mmax}=6$, $\texttt{kmax}=36$ in
\href{https://github.com/cbehan/pycftboot}{\texttt{PyCFTBoot}} and we
include spins up to $\ell_{\text{max}}=26$. For \texttt{SDPB} we use the
options \texttt{--findPrimalFeasible} and \texttt{--findDualFeasible} and
we choose $\texttt{precision}=660$, $\texttt{dualErrorThreshold}=10^{-20}$
and default values for other parameters.

\subsec{MN}
For theories with $\MN_{m,n}$ symmetry the bound on the leading scalar
singlet is the same as the bound on the leading scalar singlet of the
$O(mn)$ model. We will thus focus on bounds on the leading scalar in the
$X$ sector, which we have found to display the most interesting behavior.
Let us mention here that in the theory of $n$ decoupled $O(m)$ models the
dimension of the leading scalar in the $X$ sector is the same as the
dimension of the leading scalar in the two-index traceless-symmetric irrep
of $O(m)$.  Based on the results of~\cite{Kos:2013tga} we can see that the
theory of $n$ decoupled $O(m)$ models is located deep in the allowed region
of our corresponding $X$-bounds below.

\begin{figure}[ht]
  \centering
  \includegraphics{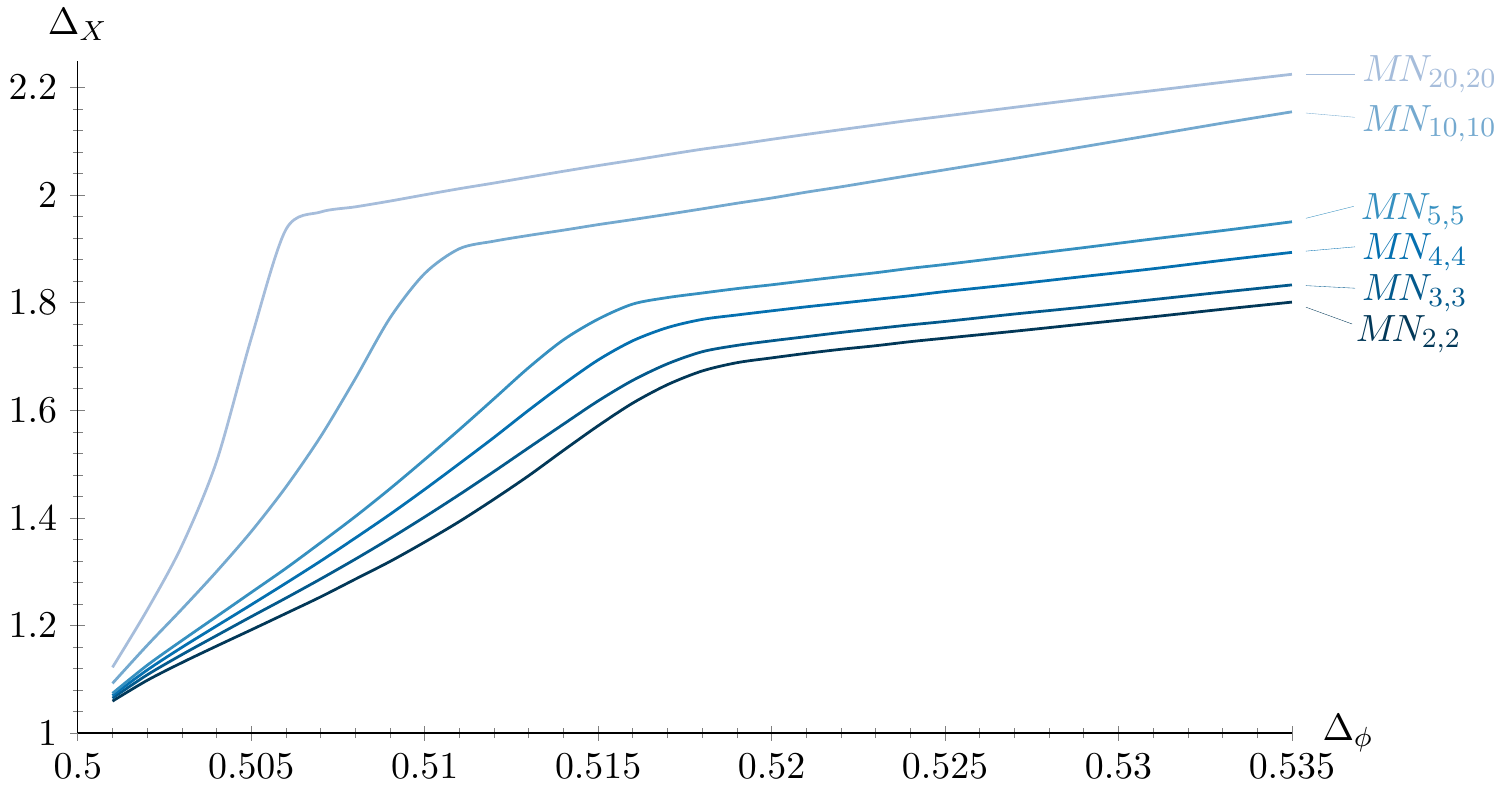}
  \caption{Upper bound on the dimension of the first scalar $X$ operator in
  the $\phi_i\times\phi_j$ OPE as a function of the dimension of $\phi$.
  Areas above the curves are excluded in the corresponding theories.}
  \label{fig:Delta_X_MN_diag}
\end{figure}
For some theories with $m=n$ the bounds are shown in
Fig.~\ref{fig:Delta_X_MN_diag}.  The form of these bounds is rather
suggestive regarding the large $m,n$ behavior of the $\MN_{m,n}$ theories.
Recall that in the $O(N)$ models as $N\to\infty$ we have
$\Delta_\phi^{\!O(N)}\to\frac12$ and $\Delta_S^{\!O(N)}\to 2$.  There is
another case where a type of large $N$ expansion exists, namely in the
$O(m)\times O(n)$ theories~\cite{PhysRevB.38.4916, Pelissetto:2001fi}.
There, for fixed $m$ one can find a well-behaved expansion at large $n$. Of
course $m$ and $n$ are interchangeable in the $O(m)\times O(n)$ example,
but in our $\MN_{m,n}$ case it is not clear if we should expect the
large-$N$ behavior to arise due to $m$ or due to $n$. It is perhaps not
surprising that it is in fact due to $m$. Keeping $m$ fixed and increasing
$n$ does not have a significant effect on the location of the kink---see
Fig.~\ref{fig:Delta_X_MN_nondiag}. On the other hand, keeping $n$ fixed and
raising $m$ causes the kink to move toward the point $(\frac12,2)$---see
Fig.~\ref{fig:Delta_X_MN_nondiag}.  (After these bootstrap results were
obtained the authors of~\cite{Osborn:2017ucf} realized that the large-$m$
expansion was easy to obtain in the $\veps$ expansion and they updated the
\href{https://arxiv.org}{\texttt{arXiv}} version of~\cite{Osborn:2017ucf}
to include the relevant formulas. The anomalous dimension of $X$ is equal
to $\veps$ at leading order in $1/m$, and so
$\Delta_X^\veps=d-2+\veps+\text{O}(\frac1m)=2+\text{O}(\frac1m)$.)
\begin{figure}[H]
  \centering
  \includegraphics{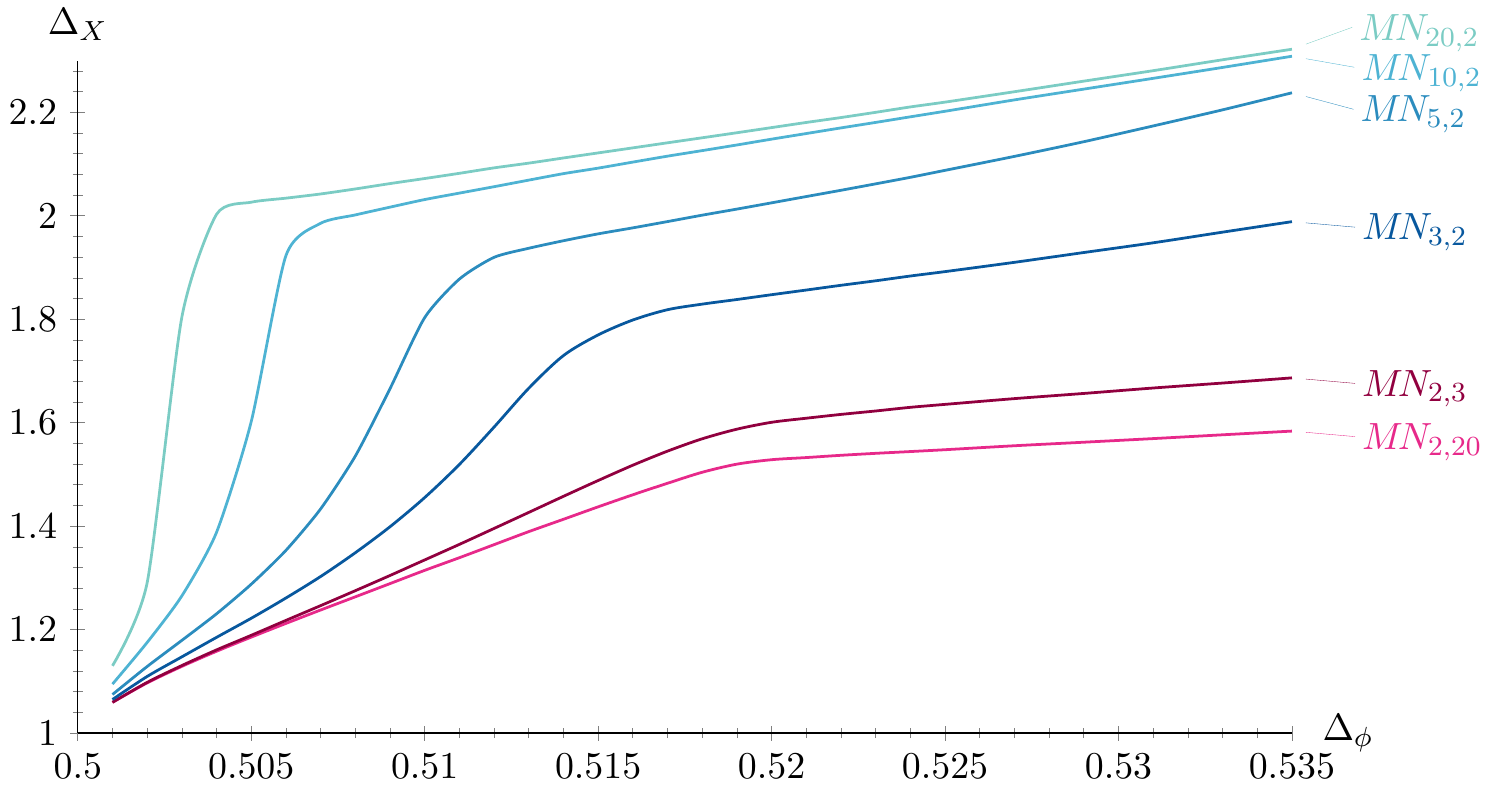}
  \caption{Upper bound on the dimension of the first scalar $X$ operator in
  the $\phi_i\times\phi_j$ OPE as a function of the dimension of $\phi$.
  Areas above the curves are excluded in the corresponding theories.}
  \label{fig:Delta_X_MN_nondiag}
\end{figure}
\begin{figure}[ht]
  \centering
  \includegraphics{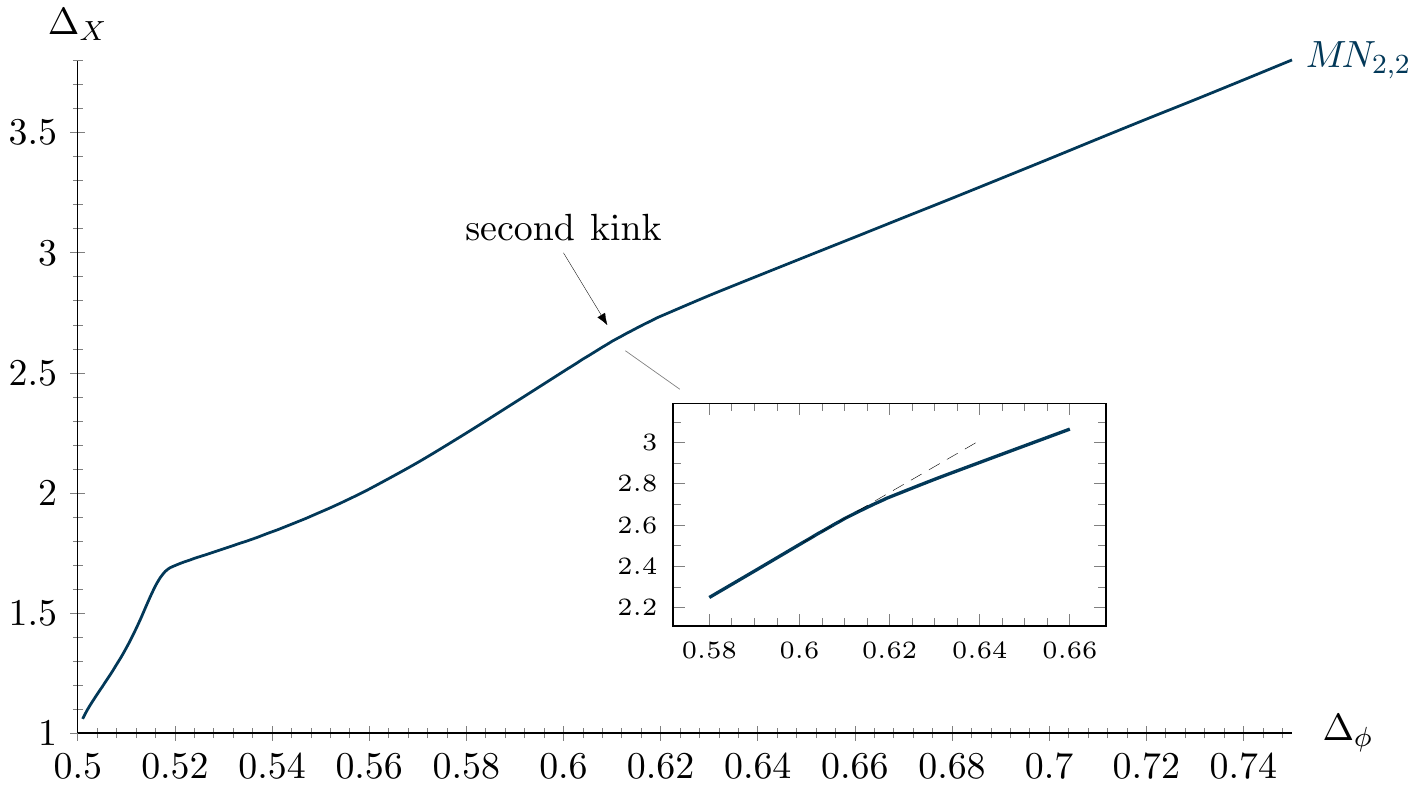}
  \caption{Upper bound on the dimension of the first scalar $X$ operator in
  the $\phi_i\times\phi_j$ OPE as a function of the dimension of $\phi$ in
  the $\MN_{2,2}$ theory.  The area above the curve is excluded.}
  \label{fig:Delta_X_MN-2-2}
\end{figure}
\begin{figure}[H]
  \centering
  \includegraphics{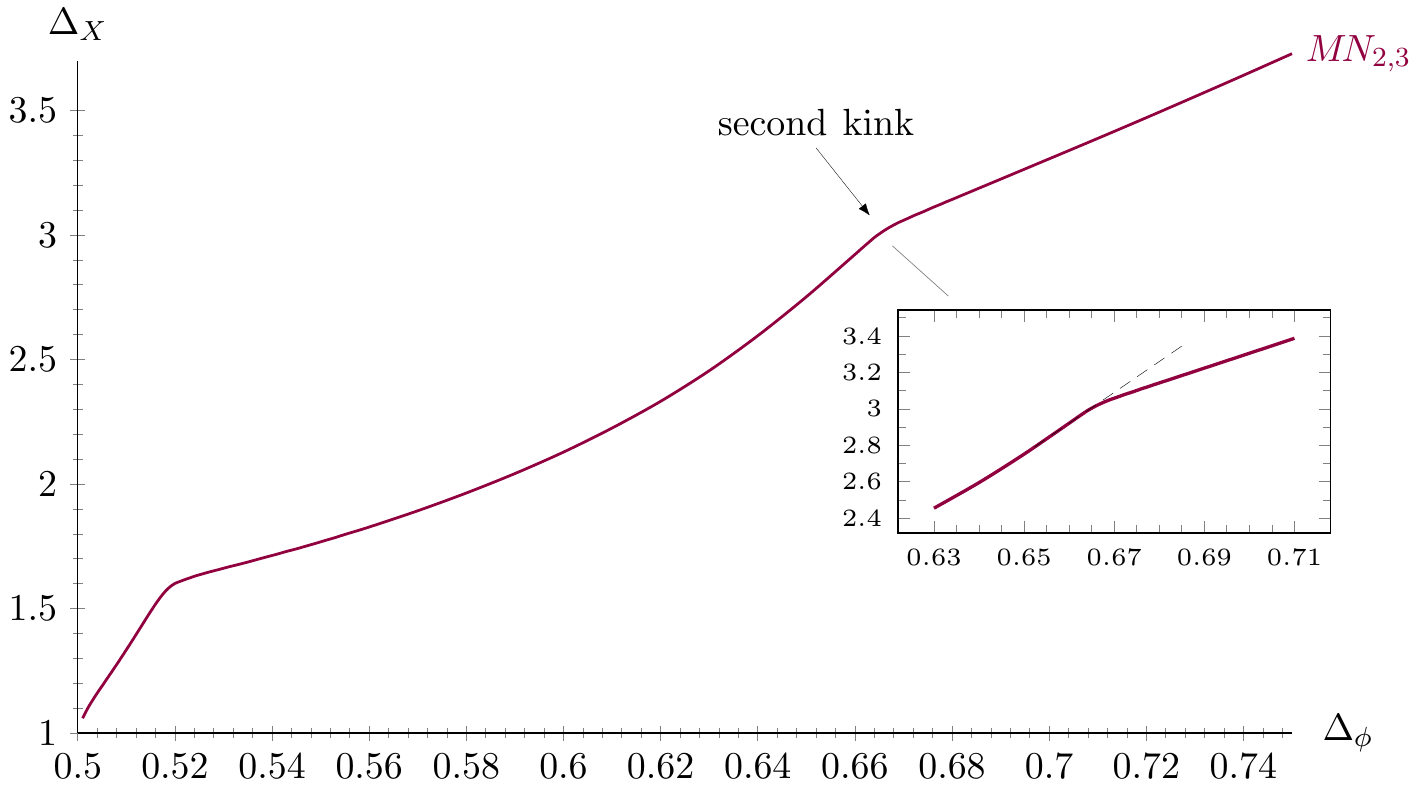}
  \caption{Upper bound on the dimension of the first scalar $X$ operator in
  the $\phi_i\times\phi_j$ OPE as a function of the dimension of $\phi$ in
  the $\MN_{2,3}$ theory.  The area above the curve is excluded.}
  \label{fig:Delta_X_MN-2-3}
\end{figure}

Continuing our investigation of the $\MN_{2,2}$ theory for larger
$\Delta_\phi$ we obtain Fig.~\ref{fig:Delta_X_MN-2-2}.  There we observe
the presence of a second kink. Although not as convincing as the kink for
smaller $\Delta_\phi$ in the same theory, it is tempting to associate this
kink with the presence of an actual CFT.  This is further supported by the
results from our spectrum analysis which give us the critical exponents
\heliCrit that match experimental results very well as mentioned in the
introduction.

Let us mention here that the spectrum analysis consists of obtaining the
functional $\vec{\alpha}$ right at the boundary of the allowed region (on
the disallowed side) and looking at its action on the vectors
$\vec{V}_{\Delta,\ell}$ of $F^{\pm}_{\Delta,\ell}$ that appear in the
crossing equation $\sum_{\text{all sectors}}
\lambda_\cO^2\vec{V}_{\Delta,\ell} =-\vec{V}_{0,0}$, where $\vec{V}_{0,0}$
is the vector associated with the identity operator. Zeroes of
$\vec{\alpha}\cdot\vec{V}_{\Delta,\ell}$ appear for $(\Delta,\ell)$'s of
operators in the spectrum of the CFT that saturates the kink and provide a
solution to the crossing equation. More details for this procedure can be
found in~\cite{ElShowk:2012hu} and \cite[Sec.\ 3.2]{Stergiou:2018gjj}. For
the determination of critical exponents we simply find the dimension that
corresponds to the first zero of $\vec{\alpha}\cdot \vec{V}_{\Delta_S,0}$.

For the $\MN_{2,3}$ theory we also find a second kink---see
Fig.~\ref{fig:Delta_X_MN-2-3}---which is more pronounced than in the
$\MN_{2,2}$ case. A spectrum analysis for the theory that lives on this
second kink yields the critical exponents \secCFT, in good agreement with
the measurement of~\cite{BakLebech}.

Another physical quantity one can study in a CFT is the central charge
$C_T$, i.e.\ the coefficient in the two-point function of the stress-energy
tensor:
\eqn{\langle T_{\mu\nu}(x)\llsp T_{\rho\sigma}(0)\rangle=
  C_T\frac{1}{S_d^{\llsp 2}}\frac{1}{(x^2)^d}\lsp
\mathcal{I}_{\mu\nu\rho\sigma}(x)\,,}[]
where $S_d=2\pi^{\frac12 d}/\Gamma(\frac12 d)$ and
\eqn{\qquad\mathcal{I}_{\mu\nu\rho\sigma}=\tfrac12(I_{\mu\rho}\lsp
I_{\nu\sigma}+I_{\mu\sigma}\lsp
I_{\nu\rho})-\frac{1}{d}\lsp\eta_{\mu\nu}\eta_{\rho\sigma}\,,
\qquad I_{\mu\nu}=\eta_{\mu\nu}-\frac{2}{x^2}\lsp x_\mu x_\nu\,.}[]
The central charge of a free scalar in $d=3$ is
$C_T^{\llsp\text{scalar}}=\frac32$.

\begin{figure}[H]
  \centering
  \includegraphics{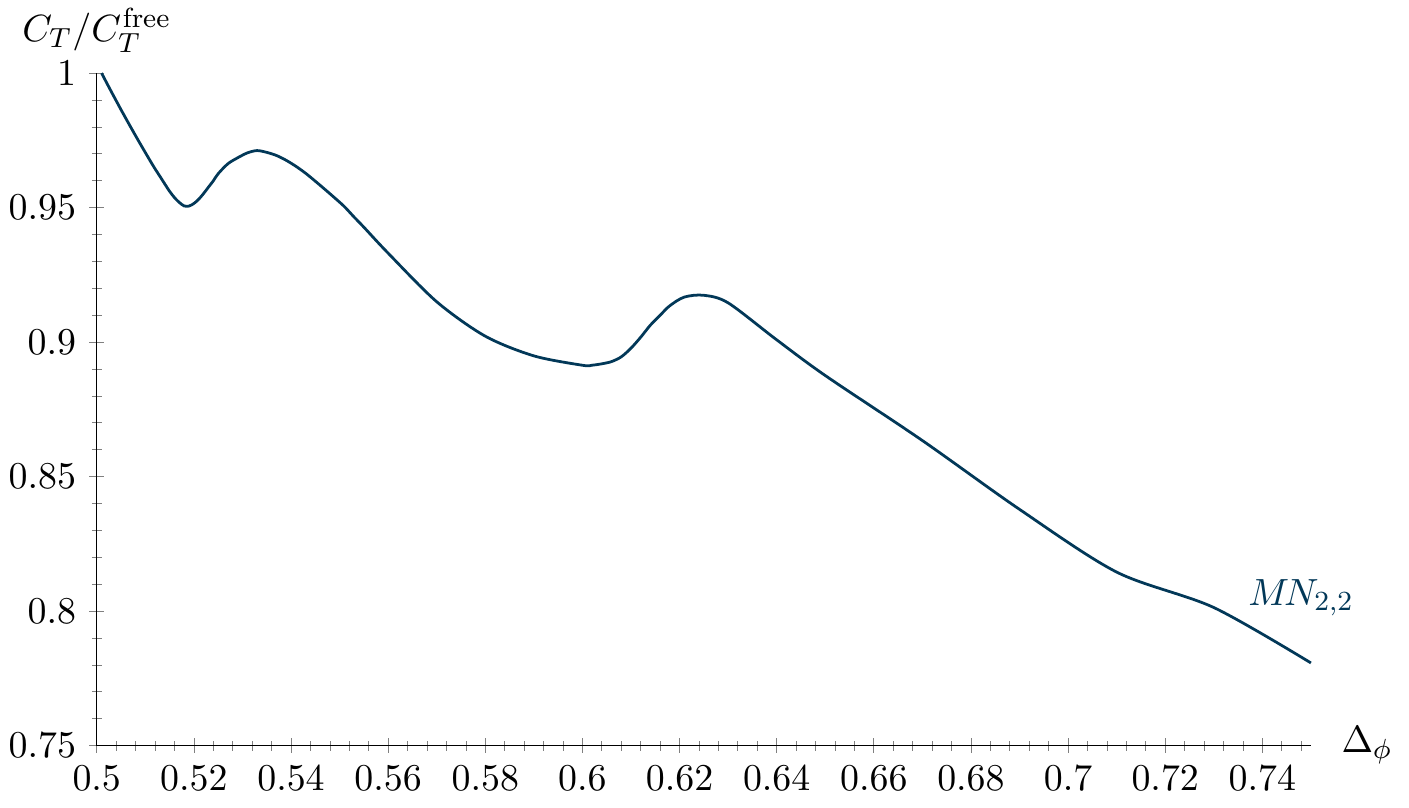}
  \caption{Central charge values in the $\MN_{2,2}$ theory assuming that
  the dimension of the leading scalar $X$ operator lies on the bound in
  Fig.~\ref{fig:Delta_X_MN-2-2}.} \label{fig:cc_MN-2-2}
\end{figure}
\begin{figure}[h]
  \centering
  \includegraphics{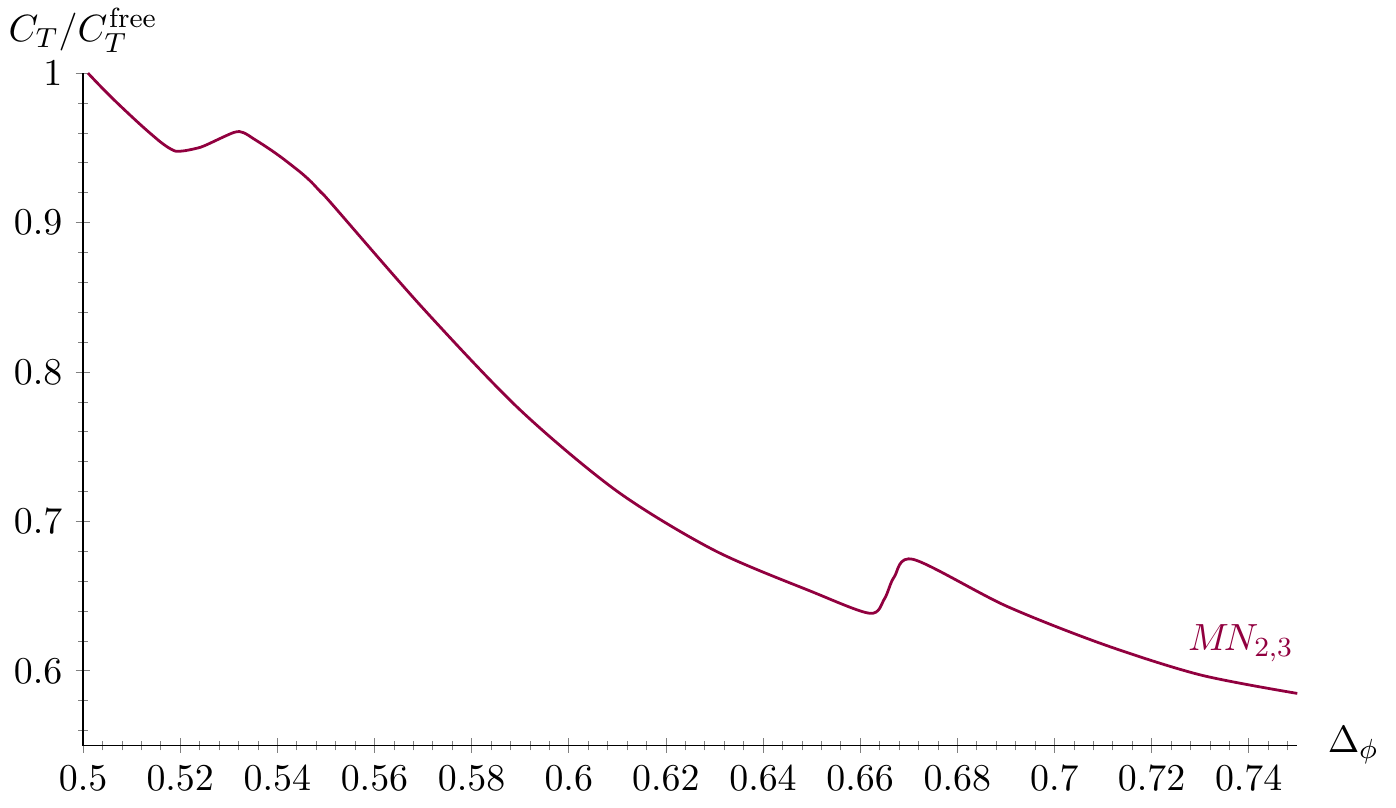}
  \caption{Central charge values in the $\MN_{2,3}$ theory assuming that
  the dimension of the leading scalar $X$ operator lies on the bound in
  Fig.~\ref{fig:Delta_X_MN-2-3}.} \label{fig:cc_MN-2-3}
\end{figure}

In Figs.~\ref{fig:cc_MN-2-2} and \ref{fig:cc_MN-2-3} we obtain values of
the central charge of $\MN_{2,2}$ and $\MN_{2,3}$ theories assuming that
the leading scalar $X$ operator lies on the bound in
Figs.~\ref{fig:Delta_X_MN-2-2} and \ref{fig:Delta_X_MN-2-3}, respectively.
The free theory of $mn$ scalars has central charge
$C_T^{\llsp\text{free}}=mn\lsp C_T^{\llsp\text{scalar}}=\frac32\llsp mn$.
We observe two local minima in Figs.~\ref{fig:cc_MN-2-2} and
\ref{fig:cc_MN-2-3}, located at $\Delta_\phi$'s very close to those of the
kinks in Figs.~\ref{fig:Delta_X_MN-2-2} and \ref{fig:Delta_X_MN-2-3}.  We
consider this a further indication of the existence of the CFTs we have
associated with the kinks in Figs.~\ref{fig:Delta_X_MN-2-2} and
\ref{fig:Delta_X_MN-2-3}.

\subsec{Tetragonal}
The bound on the leading scalar in the singlet sector in the $R_n$ theory
is identical, for the cases checked, to the bound obtained for the leading
scalar singlet in the $O(2n)$ model. The bound on the leading scalar in the
$X$ sector is identical, again for the cases checked, to the bound on the
leading scalar in the $X$ sector of the $\MN_{2,n}$ theory. Both these
symmetry enhancements are allowed, and they show that if a tetragonal CFT
exists, then its leading scalar singlet operator has dimension in the
allowed region of the bound of the leading scalar singlet in the $O(2n)$
model. A similar comment applies to the leading scalar $X$ operator and the
bound on the leading scalar $X$ operator of the $\MN_{2,n}$ theory.

\begin{figure}[ht]
  \centering
  \includegraphics{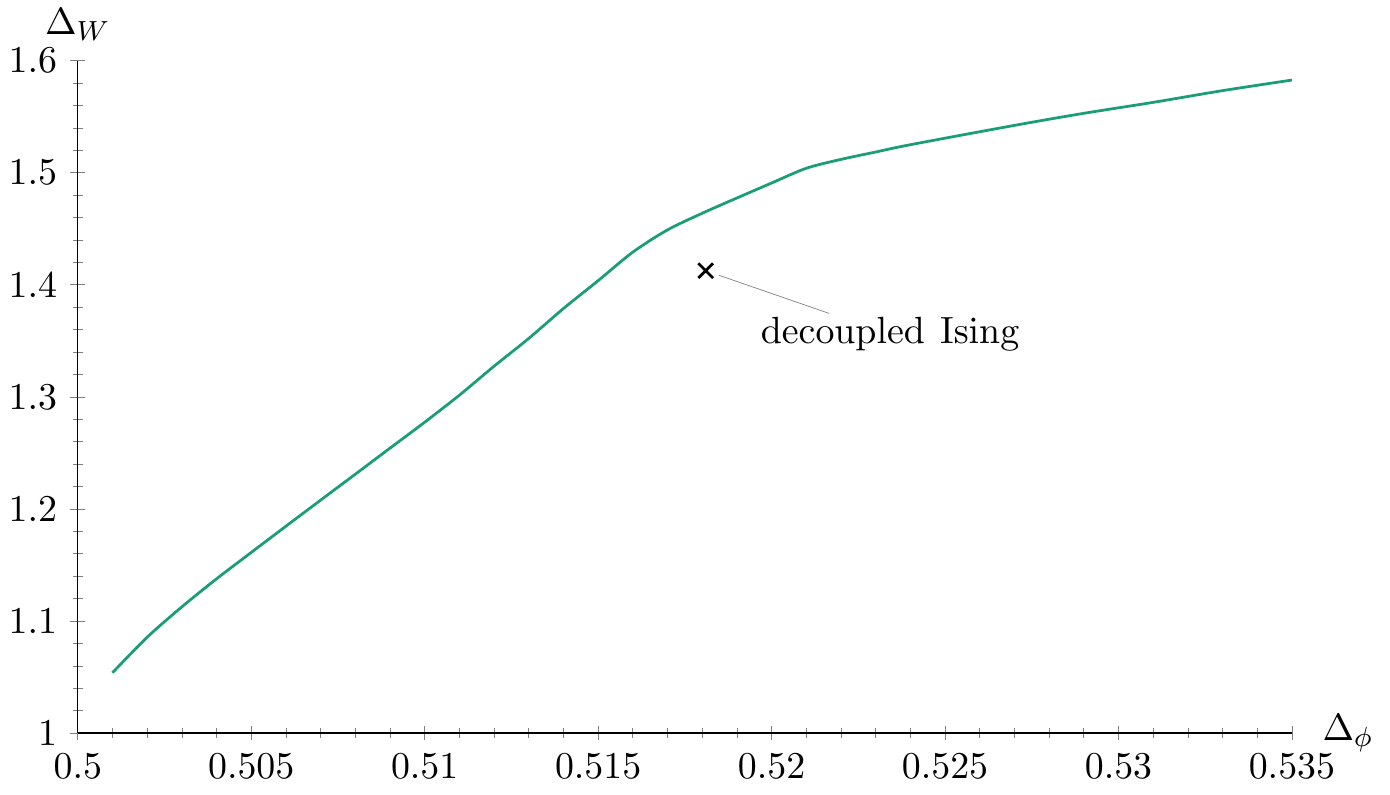}
  \caption{Upper bound on the dimension of the first scalar $W$ operator in
  the $\phi_i\times\phi_j$ OPE as a function of the dimension of $\phi$.
  The area above the curve is excluded. This bound applies to all $R_n$
  theories checked.}
  \label{fig:Delta_W_tetra}
\end{figure}

Let us focus on the bound of the leading scalar in the $W$ sector, shown in
Fig.~\ref{fig:Delta_W_tetra}.
It turns out that the $W$-bound is the same for all $n$ checked, even for
$n$ very large. It is also identical to the $V_\square$-bound
in~\cite[Fig.~14]{Stergiou:2018gjj}. The coincidence of the $W$ bound with
that of~\cite[Fig.~14]{Stergiou:2018gjj} is ultimately due to the fact that
the $N=2$ ``cubic'' theory has global symmetry $D_4$. Indeed, taking $n$
decoupled copies of the $D_4$ theory leads to a theory with symmetry $R_n$.
The leading scalar operator $V_\square$, whose dimension is bounded in the
$D_4$ theory in~\cite[Fig.~14]{Stergiou:2018gjj}, gives rise to a
direct-sum representation that is reducible under the action of $R_n$. That
representation splits into two irreps of $R_n$, namely our $W$ and $X$, and
it is easy to see that, if the $R_n$ theory is decoupled, the leading
scalar operator in the irrep $W$ must have the same dimension as
$V_\square$ of~\cite[Fig.~14]{Stergiou:2018gjj}.  Hence, the corresponding
bounds have a chance to coincide and indeed they do.  If a fully
interacting $R_n$ theory exists, then the dimension of the leading scalar
$W$ operator of that theory is in the allowed region of
Fig.~\ref{fig:Delta_W_tetra}. We point out here that the putative theory
that lives on the bound of~\cite[Fig.~14]{Stergiou:2018gjj} is not
predicted by the $\veps$ expansion.

To see if a fully-interacting $R_n$ theory exists, we have obtained bounds
for the leading scalar and spin-one operators in other sectors.
Unfortunately, our (limited) investigation has not uncovered any features
that could signify the presence of hitherto unknown CFTs with $R_n$ global
symmetry.

\newsec{Conclusion}[secConc]
In this paper we have obtained numerical bootstrap bounds for
three-dimensional CFTs with global symmetry $O(m)^n\rtimes S_n$ and
$D_4{\!}^n\rtimes S_n$, where $D_4$ is the dihedral group of eight
elements. The $O(m)^n\rtimes S_n$ case displays the most interesting
bounds. We have found clear kinks that appear to correspond to the theories
predicted by the $\veps$ expansion and have observed that the $\veps$
expansion appears to be unsuccessful in predicting the critical exponents
and other observables with satisfactory accuracy in the $\veps\to1$ limit.
However, the identification of our kinks with the $\veps$ expansion
theories is not conclusively demonstrated, especially for the interesting
cases of $O(2)^2\rtimes S_2$ and $O(2)^3\rtimes S_3$ global symmetry, and
is left for future work.

Experiments in systems that are supposed to be described by CFTs with
$O(2)^2\rtimes S_2$ symmetry have yielded two sets of critical
exponents~\cite{Delamotte:2003dw}.  Having found two kinks in a certain
bound for such CFTs, we conclude that there are two distinct universality
classes with $O(2)^2\rtimes S_2$ global symmetry.  Our critical-exponent
computations in these two different theories, given in \STAcrit and
\heliCrit, match very well the experimental results.  It would be of great
interest to examine further the conditions under which the
renormalization-group flow is driven to one or the other CFT.

For theories with $O(2)^3\rtimes S_3$ symmetry we also find two kinks. The
corresponding critical exponents are given in \Anticrit and \secCFT. The
CFT that lives on the second kink, with critical exponents \secCFT, is the
one with which we can reproduce experimental results. This is not the CFT
predicted by the $\veps$ expansion. A more complete study of the second set
of kinks that appear in our bounds would be of interest. Note that the
kinks we find do not occur in dimension bounds for singlet scalar
operators, so we consider it unlikely (although we cannot exclude it) that
the second kinks correspond to a theory with a different global symmetry
group as has been observed in a few other cases~\cite{Poland:2011ey,
Nakayama:2017vdd, Li:2018lyb}.

Beyond the examples mentioned or studied in this work, bootstrap studies of
CFTs with $O(2)\times O(N)$ and $O(3)\times O(N)$ symmetry with $N>2$ have
been performed in~\cite{Nakayama:2014lva, Nakayama:2014sba}, where evidence
for a CFT not seen in the standard $\veps$ expansion was presented.  Such
CFTs have been suggested to be absent in perturbation theory, but to arise
after resummations of perturbative beta functions.  Examples have been
discussed in $O(2)\times O(N)$ frustrated spin
systems~\cite{Pelissetto:2000ne, Calabrese:2002af, Calabrese:2004nt}. These
examples have been criticized in~\cite{Delamotte:2010ba, Delamotte:2010bb}.
However, the results of \cite{Nakayama:2014sba} for the $O(2)\times O(3)$
case are in good agreement with those of~\cite{Pelissetto:2000ne,
Calabrese:2004nt}, lending further support to the suggestion that new fixed
points actually exist.

The study of more examples with numerical conformal bootstrap techniques is
necessary in order to examine the conditions under which perturbative field
theory methods may fail to predict the presence of CFTs or in calculating
the critical exponents and other observables with accuracy. Examples of
critical points examined with the $\veps$ expansion
in~\cite{Pelissetto:2000ek, Osborn:2017ucf, Rychkov:2018vya,
Zinati:2019gct} constitute a large unexplored set. The generation of
crossing equations for a wide range of finite global symmetry groups was
recently automated~\cite{Go:2019lke}. This provides a significant reduction
of the amount work required for one to embark on new and exciting numerical
bootstrap explorations.

\ack{I am grateful to Hugh Osborn for countless important comments, remarks
and suggestions, and Slava Rychkov and Alessandro Vichi for many
illuminating discussions. I also thank Kostas Siampos for collaboration in
the initial stages of this project.  Some computations in this paper have
been performed with the help of \emph{Mathematica} and the package
\href{http://www.xact.es}{\texttt{xAct}}. The numerical computations in
this paper were run on the LXPLUS cluster at CERN.}

\bibliography{bootstrapping_MN_tetragonal}

\end{document}